Paper ID 405
# Monetisation of and Access to in-Vehicle data and resources: the 5GMETA approach

**Djibrilla Amadou Kountche[1*], Fatma Raissi[1], Mandimby Ranaivo Rakotondravelona [1], Edoardo Bonetto[2], Daniele Brevi[2], Angel Martin[3], Oihana Otaegui[3], Gorka Velez[3]**

1. {djibrilla.amadou-kountche, fatma.raissi, mandimby.ranaivo}@akka.eu, AKKA HIGH TECH, France
2. {edoardo.bonetto, daniele.brevi}@linksfoundation.com, LINKS Foundation, Italy
3. {amartin, ootaegui, gvelez}@vicomtech.org, VICOMTECH, Spain

**Abstract**

Today's vehicles are increasingly embedded with computers and sensors which produce huge amount of data. The data are exploited for internal purposes and with the development of connected infrastructures and smart cities, the vehicles interact with each other as well as with road users generating other types of data. The access to these data and in-vehicle resources and their monetisation faces many challenges which are presented in this paper. Furthermore, the most important commercial solution compared to the open and novel approach faced in the H2020 5GMETA project.

**Keywords:**
Monetisation and in-vehicle data

**Introduction**

An important ecosystem is formed by a vehicle, its users, the vehicle's manufacturer, the OEMs and, the suppliers, the service providers (insurers, repairers, etc.) and public authorities (legislators, road operators, etc.). In this ecosystem, the data generated by the vehicle and its users is considered as new gold by all the entities in the ecosystem. Therefore, the access to this data and its monetization can lead to fierce competitions between some entities in the ecosystem but the data should benefit all the actors as increasingly requested by citizens and legislators. An open, interoperable and secure vehicular platform seems the most appropriate solution for the monetization and the access to in-vehicle data and resources. Such an approach has been promoted by many European Regulations and Directives including the e-call type approval Regulation (Regulation (EU) 2015/758), the Priority area IV of Directive 2010/40/EU and the Euro 5 Regulation and Diagnostic, Repair and Maintenance Information and the upcoming Digital Services Act [8]. Such a platform poses many challenges on the legal and technical aspects which results in different solutions. The rest of the paper will introduce these legal and technical aspects, internal data sources of a modern vehicle, the different stakeholders. This next section presents the problem of in-vehicle data access and monetisation, followed by sections on the proposed solutions with an emphasis on the Data Server Platform and finally the 5GMETA platform.

**The problem of vehicle data and resources monetization**

Vehicle data and resources are at the core of a competition between various actors: car manufacturers,

Monetisation of and Access to in-Vehicle data and resources: the 5G-META approach

car dealer, etc. as depicted on Figure 1. Firstly, car owner, renter, passenger or road user might see their privacy violated by unlawful data collection, their rights to repair improved or limited by the solutions adopted by others actors for the access to these data; The OEMs and other services providers (Repair and Maintenance, vehicle oil suppliers, etc.) might also offer competing services to the citizen. Therefore, the challenges facing any solution are to guarantee fair access and competition as well as right to repair and privacy.

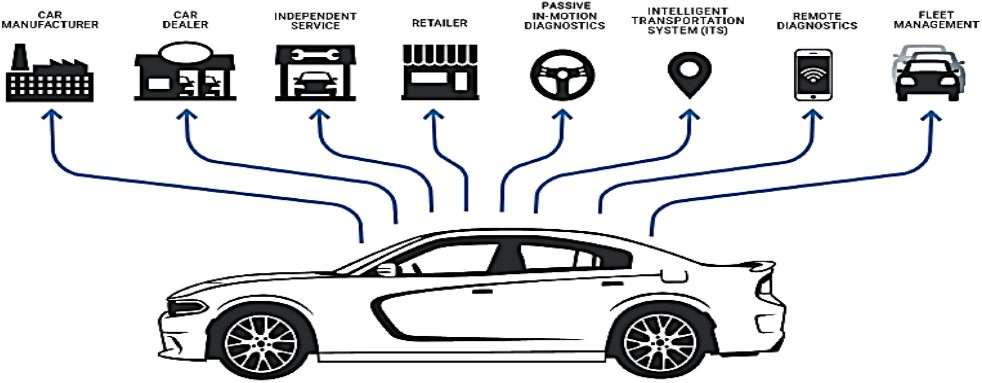

**Figure 1 Illustration of the different actors [9].**

Several technologies are used to access vehicle data and store them in the cloud/edge of the network, among them the most important are: i) The CAN bus that is the de-facto standard used by OEMs to exchange data and connect sensors inside the vehicle. Typically, the data format is proprietary and different for each OEM, moreover, in the last years, some OEMs started protecting data using encryption. ii) The On-board Diagnostic II (ODB II) provides access to emission and other data exchanged on the CAN bus and allows professional repairers to assess the status of the vehicle. In the last years, many products on the market exploit the ODB II interface to create a remote access to these data. iii) The Fleet Monitoring Standard [10] allows the access to trucks and buses data exchanged on the CAN bus. An evolution of the standard called Remote Fleet Monitoring (rFMS) [11] allows access to these data by using the manufacturer backend servers.

With the advent of C-ITS/CCAM concepts, cars start exchanging data using different types of communication channels such as short range (WAVE/ITS-G5 and C-V2X) and long range 4G/5G (through the Uu interface). The exchanged data is of different types like e.g., status data (ETSI CAM, DENM, etc.) or more complex information to build cooperative message (for *e.g.,* CPM, MCS,). The information used to feed these messages are typically read directly by the in-vehicle CAN network.

While in short-range data are directly exchanged among vehicles, cellular connectivity allows information to be collected and stored in the network. The advent of 5G will increase the possibility to collect data thanks to URLLC, eMBB and mMTC profiles that provide low-latency, high bandwidth and scalability unlocking several applications that were not feasible with 4G networks.

All this information is nowadays collected in several ways e.g., using OEMs' proprietary cloud servers with different aims like maintenance. A standardization effort in this direction is the Extended Vehicle concept by ISO (ISO 20077-20078-20080) where data servers and interfaces are described with a strong focus on privacy, data protection, and cybersecurity. Data is also collected by infotainment services again using manufacturer's proprietary server or, in the case of smartphones by the corresponding applications providers like Google, Apple, etc.

With such a jeopardized situation, the access to data is subject of debate for the lawmaker: a recent legislation passed in Massachusetts (**Massachusetts Right to Repair Initiative (2020)**) require all vehicle sold in the state to have a standardised open access data platform which will allow with the permission of the car's owner to provide access to all vehicle data including telematics, navigation data (GPS) and mobile Internet to repairers and mechanics. Without this legislation, the access to these data can only be done through the manufacturer's shops with their permission [12, 13, 14]. In the EU, many regulations have been passed to address this issue. These regulations mandate the creation of an open, interoperable and a secure platform which follow these principles [2]: Data provision conditions: Consent, Fair and undistorted competition, Data privacy and data protection, Tamper-proof access and liability, and Data economy. Therefore, the management and monetisation of all this data is still an open topic that needs to be addressed.

**Proposed Solutions**

Different solutions have been identified and discussed by the C-ITS platform [2]: the Data server platform, the in-vehicle interfaces and the on-board application platform in order to fulfil the legal requirements for an open, interoperable and secure platform. Each of these solutions is discussed in a separated section. 5GMETA is funded by the EC (Grant 957360) and aims to have a direct impact on the value chain and the business models that will be analysed within the project:

- Generation of new business models: new value creation models for data generators and consumers generating revenues from direct monetization of innovative services or applications;
- Product optimization: Automotive players would reduce costs by means of shorter and efficient iterations on design and training from data coming from real-field tests prototypes or mass-produced systems in real mobility environments;
- Increasing safety and security: With other users and public entities getting triggers on time to reduce time for intervention.

The following sections will describe the technical solutions and their implementation when available on the market. The emphasis will be set on the data server platform which is provided by Alibaba, Amazon, Google, and Microsoft. The 5G-META solution will also be extensively described.

**The data server platform**

The data server platform represents a generalisation of the architectures already used in C-ITS, CCAM, and in many research projects (e.g., H2020 projects) where central servers receive data from the vehicle and stored it in databases to provide different sets of services. This solution is relatively easy to deploy and is the solution of choice of the vehicle manufacturers. Also, many Cloud providers offer their solutions tailored to the automotive sectors that will be described below. There are three version of the data server platform: the extended vehicle concept, the shared server and the Marketplace [2].

*Alibaba Automotive Logistics Solution*

Alibaba's Automotive Logistics (AAL) Solution is part of Alibaba's IoT solutions focusing on vehicle monitoring and management. It is then primarily intended for vehicle manufacturers and goods owners who are interested in vehicle monitoring, fleet management but also other services in the domain of logistics. The solution collects and processes the vehicles data along with their location. Based on the constituted datasets, it offers the users of the solution (fleet manager, truck drivers, cargo owners, etc.) big data services, answering the queries received from the users' applications (e.g., monitoring applications). The solution allows also advanced data analysis supported by available analytical models. AAL's architecture consists of 4 zones as illustrated on Figure 2**Erreur ! Source du renvoi introuvable.**.

- IoT zone: it consists in a data collection system where the vehicle data is collected and feed to the Big data zone.
- Big data zone: it can be considered as a common pool supported by big data tools allowing the creation of multiple datasets and facilitating information flow between multiple actors (fleet manager, good owners …).
- Service zone: it is where data access and monitoring services are made available to the interested applications.
- Business intelligence zone: it hosts multiple data analysis models offering advanced data analysis services such as data statistics analysis or other business-related services.

*Amazon AWS Connected Vehicle Solution (AWS CVS)*

This solution allows automotive players to *gather, analyse and act on vehicle data without having to manage any infrastructure.* The AWS CVS does not provide a Road infrastructure component as illustrate on Figure 3**Erreur ! Source du renvoi introuvable.**. The AWS IOT Greengrass Core is used to deploy IoT applications capable of local processing, messaging, data management and machine learning inference. Thus, IoT application can be deployed using Docker container, AWS Lambda and local. AWS Greengrass supports messaging protocols such as MQTT to send data to the AWS Cloud through the AWS IoT Core and can interact with any IoT device using FreeRTOS or AWS IoT Device SDK. Upon reception of messages, the AWS IoT authenticates, authorizes and routes them to the different applications. These applications provide services such as anomaly detection, trip data, driver safety score, diagnostic trouble code, location-based marketing and notification service. The AWS CVS has already been used by BMW Group, Uber, Wireless Car and Toyota.

Monetisation of and Access to in-Vehicle data and resources: the 5G-META approach

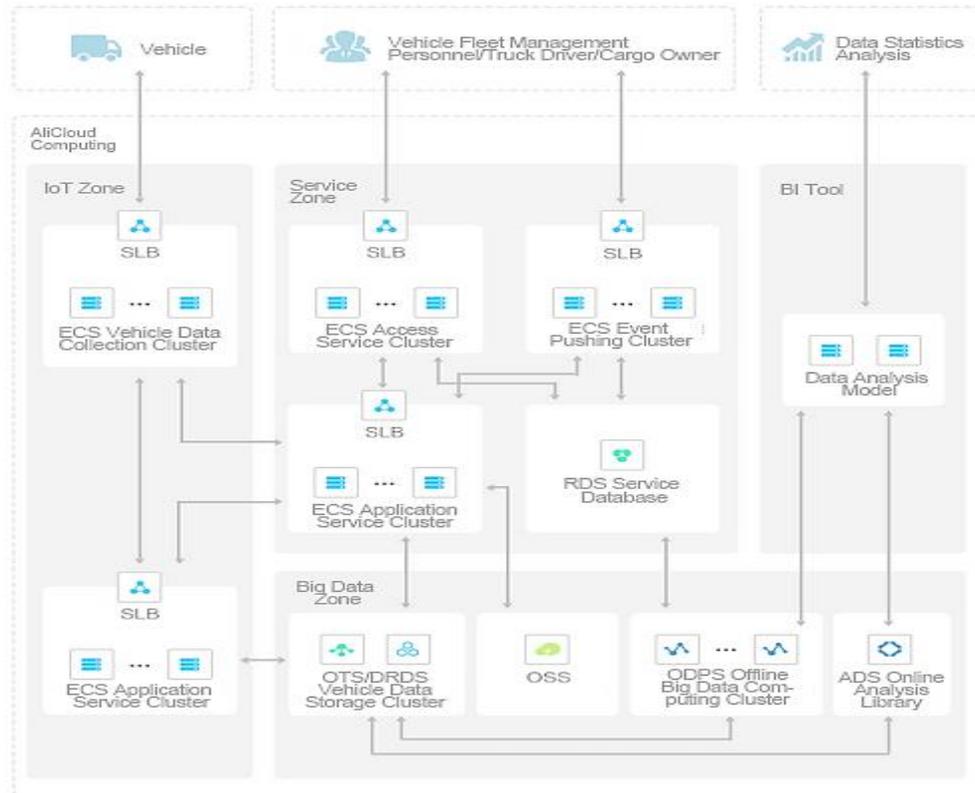

Figure 2 Alibaba Automotive Logistics Solution

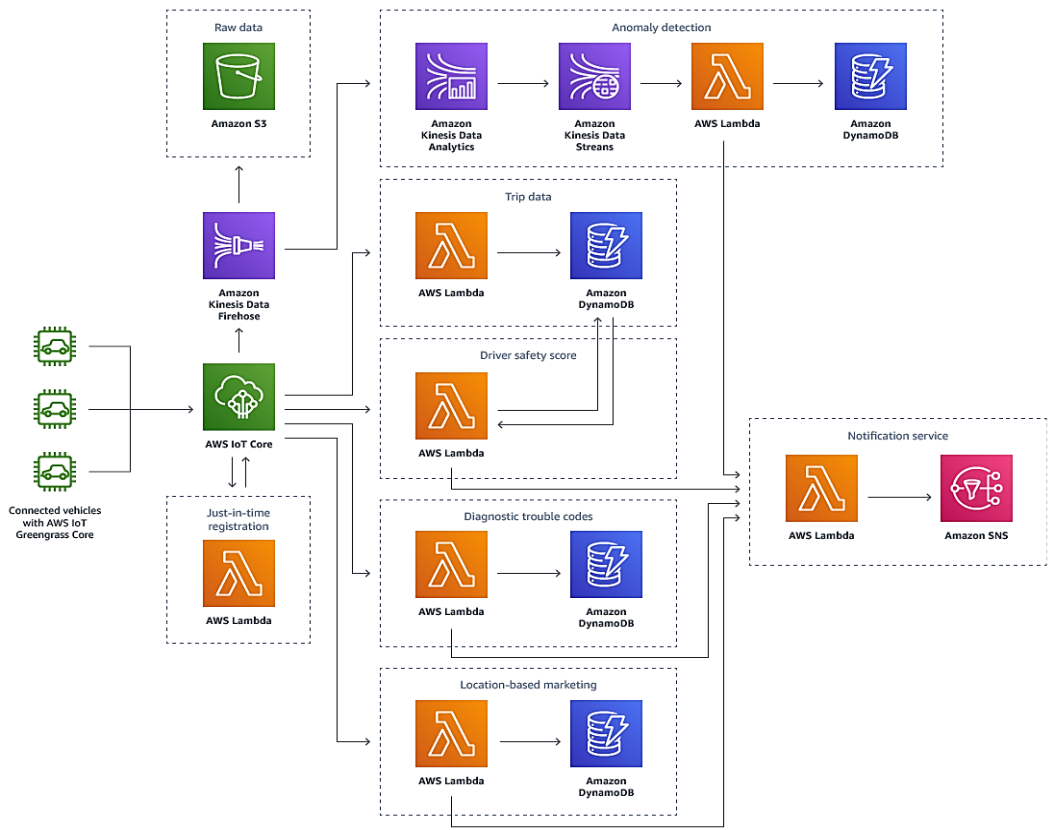

Figure 3 Amazon AWS Connected Vehicle Solution

Monetisation of and Access to in-Vehicle data and resources: the 5G-META approach

*Google* IoT Core

Google introduces the Cloud IoT core as the gateway to the Google Cloud Platform (GCP), a tool that is built to expand the capabilities of all smart devices among which connected and autonomous vehicle. The latter generates an important amount of data that should be handled by a capable cloud and edge platform. Google gives the opportunity to design this Connected Vehicle Platform (CVP) through the Cloud IoT Core and enable a plethora of use cases one of which is usage-based insurance. Google provides an architecture diagram to shape the usage-based insurance use case (see Figure 4**Erreur ! Source du renvoi introuvable.**)[15].

This application evaluates the vehicle data to determine which policy pricing should be applied to the customer. The architecture diagram depicts the two Cloud IoT Core units, namely the Device manager (1) and the MQTT Protocol Bridge. The Cloud IoT Core hosts different registries of different devices namely connected vehicles and allow them to drive their data to the GCP, where different process services are available as well as the needed security to make these communications [6].

Google IoT Core made it possible for other companies to build a CVP. For instance, Derive Systems [16], a software company, realized that although vehicles nowadays are extremely capable of customization, different car manufacturers still build vehicles using a One-Size-Fits-All software regardless of their usage (e.g., package delivery, domestic chores, law enforcement). Thanks to their CVP, they are able of gathering insights by analysing different data usage such as fuel consumption and speed. These insights helped their software engineers to develop a software upgrade that embodies the vehicle's own use.

*Volkswagen Automotive Cloud on Microsoft Azure*

Volkswagen Automotive Cloud is a dedicated automotive industry cloud which relies on Microsoft Azure technology. The project has been officially launched in 2018 and is still under development at the time of writing. It will form a bridge between the connected vehicles and the cloud-based platforms allowing the creation of various services in the area of mobility, user experience, vehicle upgrades. It is expected to handle data from millions of vehicles per day with the first customer-ready services to be rolled-out in 2022[17]. No detailed technical architecture is yet available, but an illustration is given by Figure 5**Erreur ! Source du renvoi introuvable.**.

**In-Vehicle interfaces**

Although widely used in current vehicles to easily access standardised sets of data, the OBD-II interface lacks sufficient security mechanisms to protect the vehicle from attacks [3]. Thus, new in-vehicle interfaces are needed to allow streams of data from the vehicle to be made available through wire or wireless communication (ODB+, Wi-Fi, 4G, and 5G) to consumers. This solution requires strong security and privacy mechanisms to guarantee data owners right, secure and encrypted communications and a comprehensive security strategy for the physical layer and the firewall for the different buses [3].

Monetisation of and Access to in-Vehicle data and resources: the 5G-META approach

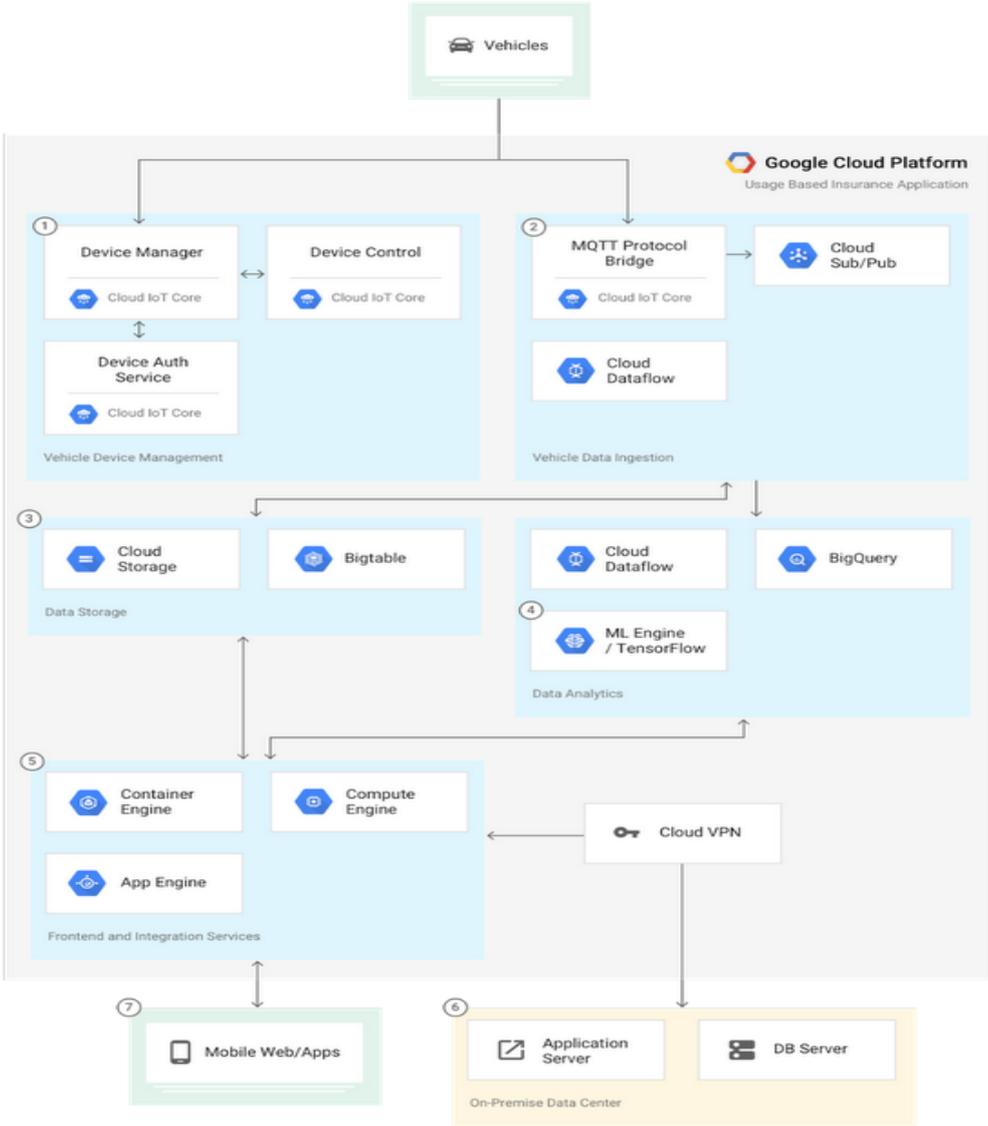

**Figure 4 Google Connected Vehicle Platform on Cloud IoT Core**

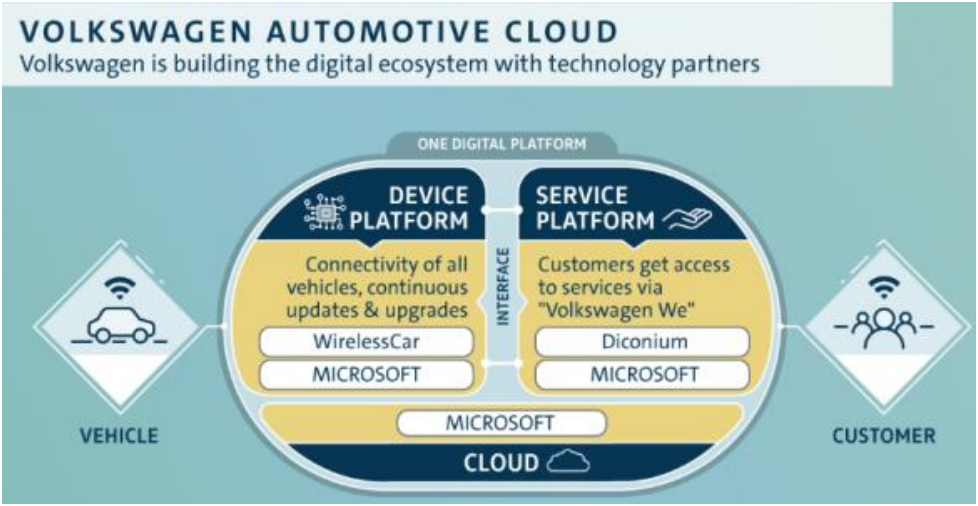

**Figure 5 Volkswagen Automotive Cloud on Microsoft Azure**

**On-board application platform**

The On-Board Unit (OBU) can soon become the hub to collect data coming from all over the vehicle, dispatching them mainly on the surrounding (via ETSI ITS G5/C-V2x or 4G/5G Uu through MEC) or towards a central collector that can be placed at the edge of the network (MEC approach) and/or in the cloud. The 5GMETA approach will be based on a flexible and modular OBU to ease the collection (and the subsequent monetization) of data.

The networking arena is under strong evolution thanks to new technologies like SDN and NVF that will evolve the network to be programmable, flexible, and scalable in a completely automated fashion. This objective will be also achieved thanks to orchestration and AI that will permit fine control of all aspects of automatic resource allocation even reacting in real-time to changes in the network status. Modern OBUs can be part of this paradigm change, becoming an active part of the network. New OBUs can dynamically host Virtual Network Functions (VNF) gaining flexibility but also on easier update procedures. In this way the OBU can, for example, host different modules to exchange data adapting to new standards, manage several technologies at the same time, and being flexible to host new approaches.

**The 5GMETA approach**

With greater proliferation of connected and automated mobility applications, the value of data from vehicles is getting strategic not just for the automotive industry but in a wider scope, and not limited to the on-board systems and services. 5GMETA open platform aims to leverage car-captured data to stimulate, facilitate and feed with them innovative products and services. Inspired by activities and results from international bodies participated by the partners (5G-PPP, 5GAA, ETSI/3GPP), 5GMETA leverages cutting-edge but internationally accepted solutions (C-V2X, MEC, new CCAM services) to build a secure and private pipeline that manages data computing, data flows according to service subscriptions and geographic queries endowed with the following features - as depicted in Figure 6:

- Secure and private mass distribution of data from vehicles: The automotive players, the OEM/TIER1, SMEs and high-tech start-ups will be provided with the car data to feed data-based innovative connected and automated mobility services and applications granting access to car data ensuring privacy/anonymization and secure/encrypted access by design.
- Data ownership: Keep data source records and subscription accountability to control the data license according to monetization models (free to use, shared revenue, freemium,), service purposes (public, private, non-profit) time-to-live (live processing=no storage permitted, derivate data storage, raw data storage,) and geographic limits. The producer (OEM, TIER1, driver, passengers) can have full control the data access and utilization by configuring the type of applications which could have access and the business model which applies licensing the data flows. This granular access register can lead to diversify the market and improve the return on investment for producers by getting other streams refer to and re-use the content.
- Scalable management: Enable a modular solution to book assets at the 5G network edge to process the incoming data flows. Available Service Level Agreements (SLAs) fitting to a well-

- balanced data throughputs and processing infrastructure costs trade-off.
- Geo-based range: Increase the flexibility of services and applications to be able to consume the data from sources at queried locations, spots, local areas, cities, or countries. This configuration over specific areas reduces management and processing needs allowing services to focus on relevant data.
- Data interoperability: Make heterogeneous data and locations homogeneous in terms of structure, schema and timestamps to ease a universal processing of CAM services and applications, while enabling and integration on top of existing standards and extending them, when necessary, and providing open common interfaces and agreed APIs.
- Real-time data messaging: Specific IoT messaging frameworks will be necessary to develop a global view across the different infrastructures, cars, network and services co-operating in CAM applications.
- Flexible business-driven configuration: Provide dashboards to configure service/application subscriptions to data catalogues, together with the integration of APIs to dynamically setup sample rates, establish target geolocations and book processing assets. The overall accountability of consumed data is provided for billing and forensics data liability, as well as additional performance reports to support decision making and troubleshooting.

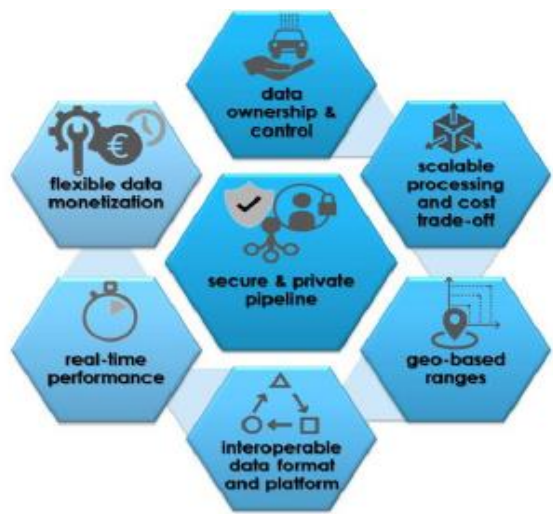

**Figure 6 Basic features of 5GMETA**

*Design principles of the 5GMETA solution*

The 5GMETA architecture is designed as open architecture to allow access to in-vehicle data and resources and allows the actors in the sector to add upgrade and swap the architecture components. Therefore, the specification of 5G-META architecture component will be published. This open architecture aims to allow the development of data markets and compliance to EU laws requesting an open architecture when accessing in-vehicle data. Thus, this architecture allows interoperability, portability and use of open standards to avoid vendor lock-in [7]. This 5GMETA open architecture embeds many components to ensure security and privacy by design and by default. In this line, edge computing offers a more efficient and secure alternative where data is processed and analysed closer to the generation point. Furthermore, as the data volumes scales-up a centralized infrastructure is not effective, and a decentralized option meet centralized drawbacks. Specifically, edge computing gains relevance as it can focus on a local area with zero latency and enforcing data privacy. Interoperability is also an important design principle of the 5GMETA architecture and will be tested in the different use cases by 5GMETA consortium and Start-ups. Finally, the 5GMETA solution will be manageable by Infrastructure as Code to easily deploy, configure, provision and orchestrate.

*The 5GMETA solution*

The open architecture, depicted by Figure 7, is composed by the 5GMETA Cloud Platform, the 5G Network, the Road Infrastructure and the Sensor and Devices components.

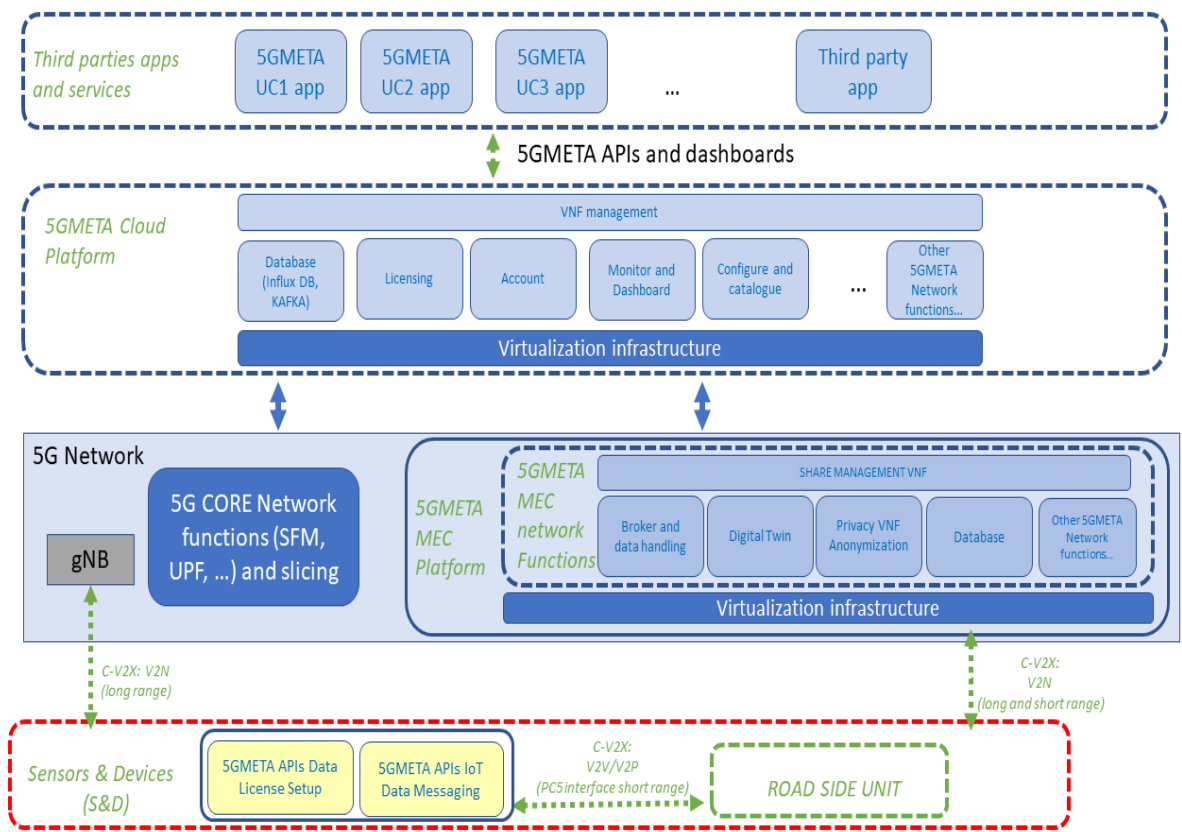

**Figure 7 5GMETA architecture**

The 5GMETA cloud platform aims to manage geographic and time queries to feed subscribed services and applications with live data, a register with the data consumed to support the billing system, a dashboard for value chain players to monitor data throughput statistics and SLAs, APIs to configure the relevant data to be subscribed, the sampling rate and to book platform assets, the credentials to access to encrypted data flows and the navigation through data inventory, and an access mechanism to limit and control the access to data flows according to applicable terms and conditions from data owners and producers.

The 5G network infrastructure component is responsible of interfacing with 5G networks in order to gather the data needed by a service. It exploits the 5G NR. 5GMETA Road Infrastructure component plays a similar role as the 5G network component in terms of the data needed at the road infrastructure. The Sensors and Devices component is deployed in the vehicles and allows secure access to in-vehicle data through messaging such as MQTT and a standardised API.

From the 5GMETA point of view, the OBU could host different VNFs devoted to the collection, formatting, and delivery of data towards the 5GMETA platform. In this way, 5GMETA can support different ways of transporting data (even in parallel), giving the application developers, that will work on the edge/cloud data, a unique vision of the collected information, without caring about how they are transported, formatted, and protected on the channel. Moreover, this leaves the OBU provider free to use their own strategy keeping the compatibility with the 5GMETA platform that has only to implement the corresponding module. Eventually, it is worth considering that Cybersecurity is a critical aspect of the 5G-META platform where different mechanisms (like DTL and TLS1.3) will be used to protect the data, the VNF and the communications. From the anonymization point of view, while is some situation data can already be anonymous, 5GMETA will provide VNFs devoted to data anonymization in a GDPR compliant way.

**Conclusion**

This paper presents the problems and challenges related to the monetization and the access to in-vehicle data. Due to the importance of vehicle data to future markets and services and the completion among the actors of the automotive industries, the ICT industry and the citizen, an open, interoperable and secure platform need to be provided in order to allow new markets. These solutions are the data server platform, the in-vehicle interface and the on-board application platform. During the 5G-META project a new platform will be implemented and tested in order to enable such a market.

**References**[1]

---

[1] The URLs were all accessed the 23rd of July 2021.